\documentclass[twocolumn,
preprintnumbers,nofootinbib,aps,prd,floatfix]{revtex4}
\pdfoutput=1 

\usepackage{subfigure,graphicx,amsmath,amssymb,hyperref}
\usepackage{youngtab}

\newcommand{\bea}{\begin{eqnarray}}
\newcommand{\eea}{\end{eqnarray}}

\def\beq#1\eeq{\begin{align}#1\end{align}}
\def\beqnn#1\eeq{\begin{align*}#1\end{align*}}

\usepackage[usenames,dvipsnames]{xcolor}
\definecolor{darkgreen}{rgb}{0,0.5,0}

\begin{document}
\preprint{UCI-HEP-TR 2020-06}
\title{On the Calculation of Invariant Tensors in Gauge Theories}

\author{Yahya Almumin}
\email{yalmumin@uci.edu}
\author{Jason Baretz}
\email{jbaretz@uci.edu}
\author{Arvind Rajaraman}
\email{arajaram@uci.edu}
\affiliation{Department of Physics and Astronomy,\\
University of California, Irvine, CA 92697-4575 USA
}

\begin{abstract}
We present an efficient method for finding the independent 
invariant tensors of a gauge theory.
Our method uses a theorem relating  invariant tensors and 
D-flat directions in field space.
We apply our method to several examples-- 
$SO(3)$ with symmetric tensors, $SU(2)$ with a dimension 4 representation, and $SU(3)$ with matter in the sextet-- and 
find the set of independent  invariant tensors in these theories.
\end{abstract}

\maketitle


A gauge theory is specified by the gauge group and 
the representation of the matter fields
under the gauge group,  
 but all observables, including the physical spectrum, 
 are gauge invariant combinations of fields. The structure of
 these objects is found by contracting the
 gauge-covariant fields with 
 {\it invariant tensors} to form invariant objects.
While there are an infinite number of invariants, 
there is expected to be a basis set of invariants such that all other invariants can be generated from these basis invariants.
 These motivate us to understand the basis set of invariant tensors for a general gauge theory.

Invariant tensors are known for the fundamental representations of the 
classical groups\footnote{For some of the mathematical 
literature relevant to invariant theory see:~\cite{Weyl:1946}--
\cite{Gufan:2001}.}. However, the tensors for
many other representations are not  classified. For
many  groups such as $E_6$, even the full set of tensors
for the fundamental representation have not been found~\cite{Pouliot:2001iw}.

A brief example will suffice to show the kinds of difficulties that may occur.
It is known that in the group $SO(3)$, the invariant tensors are $\delta^{ij}, \epsilon^{ijk}$. These
tell us that in a theory where all fields $V^I_i$ (we are using lower case letters for the gauge representation, and upper case to label the fields--a flavor index) are in the fundamental, the complete set of
invariant polynomials are generated by
$V^I_iV^J_j\delta^{ij}$ and
$V^I_iV^J_jV^K_k\epsilon^{ijk}$.
But in a theory with $SO(3)$ gauge symmetry and
with fields 
 $V^I_{ij}$ in the symmetric tensor representation, one can produce an infinite set of
 invariants by contracting
an arbitrarily long sequence $V^{I_1}_{ij}V^{I_2}_{jk}..V^{I_M}_{li}$ (and there exist
further invariants involving epsilon tensors).  Only a small set of these are independent, but finding these is nontrivial.

In this paper, we present a new  approach to
finding 
a minimal set of  
invariants for more complicated gauge theories.
Our approach will be
to use the connection between invariant tensors and 
D-flat directions in field space, which was
originally described for supersymmetric field theories
(specifically in the context of dualities in these theories \cite{Seiberg:1995pq,Kutasov:1996ss,Brax:1999gy,Pouliot:1999yv}.)
This theorem 
 asserts that  the independent gauge invariant polynomial invariants of the theory
are in  1-1 correspondence
with the orbits of 
constant  field configurations
where configurations differing by
a {\it complex} gauge transformation
are identified 
\cite{Buccella:1982nx,Procesi:1985hr,Gatto:1987bt,Gherghetta:1996dv,Brax:2001an,Luty:1996sd}.
It is clear that at any point on configuration
 space, we can calculate the value of
any gauge invariant combination of the
fields. 
The theorem states that this can be reversed; a knowledge of the values of all the independent 
gauge invariant polynomials is sufficient
to reconstruct the orbits of constant field configurations quotiented by complex gauge transformations.

This theorem is often used in simple 
theories 
to characterize the field space in terms of the known
operators.
 Here we will reverse the implication, and 
use the field space
to find a complete set of gauge invariant 
objects in various theories. 

The procedure for finding the tensors is then as follows. We take a set of 
fields in the relevant representation, and set each component to an arbitrary constant value. We then use a complex gauge transformation to set some field components to zero.
If the gauge transformations are completely fixed by this procedure, then the nonzero components parametrize the orbits, and we must find a set of invariants such that each of these parameters can be written as a linear combination of invariants. Such a set of invariants would then be a basis set of invariants for the theory.

In practice, we find that the full gauge symmetry is not easy to fix with a single field. In each case, we find a remnant discrete symmetry, and occasionally, a larger continuous symmetry. 
One possibility is to use further fields
to completely gauge fix the symmetry, but in each case we analyze below, we find that the residual symmetry is simple enough that we can find the complete set of  combinations which are invariant under the residual symmetry.
These combinations parametrize the gauge fixed space, and we must find a set of invariants such that each of these combinations can be written as a linear combination of invariants. Such a set of invariants would then be a basis set of invariants for the theory.

We now show the practicality of this approach
by explicitly finding the basis set of  invariants for
three gauge theories --$SO(3)$ with symmetric tensor matter, $SU(2)$ with matter in the dimension 4 representation, and $SU(3)$ with matter in the sextet. To our knowledge, the last two are completely new analyses (the first case has been analyzed previously in \cite{Berger:2018dxg}). 

\section{SU(2) with fields in the dimension 4 representation}

 We will take as our first example a 
 theory with a gauge group
 $SU(2)$ and a field content 
 where there are $N$ fields in a representation of
dimension 4; this is the simplest case for which the
independent set of invariants has (to our knowledge)
not been worked out.

The fields can be represented as  three-index tensors
$V^{I}_{abc}$ where $a,b,c=1,2$ are acted on by the gauge symmetry,  and $I=1..N$ labels the
different fields (we shall consistently  use lower
case indices for gauge indices and upper case indices to label
the different fields, similar to a flavor index).
The fields  can also be represented as a column 
vector with four elements:
\bea
V^{I}=\left(\begin{array}{c} V^{I}_{111}\\V^{I}_{112}\\ V^{I}_{122}
\\ V^{I}_{222}\end{array}\right)
\eea

The invariant tensor is $\epsilon^{ab}$, but
one can write an infinite set of invariants, and
it is hard to find relations between them.
We therefore find the gauge-fixed configuration space, and attempt 
to 
characterize this space by invariants.

\vskip 1 cm

We begin by considering a single field $V^{1}$. By 
a complex gauge transformation, one can set the second and third components to zero, and  set the first component 
to 1. The field then has  the form
\bea
V^{1}=\left(\begin{array}{c}1\\
0\\0\\d\end{array}\right)
\label{fieldform1}
\eea 

This breaks the continuous gauge symmetry but preserves a discrete symmetry, which can be understood as follows: if we
interchange every gauge 1 index with a 2 index, this is equivalent to taking $\epsilon^{ab}\to -\epsilon^{ab}$.  
Then any invariant with
$4n+2$ fields will pick up a minus sign, while any invariant with
$4n$ fields is unchanged. This then indicates that
the combined transformation 
of
interchanging every 1 index with a 2 index,
and multiplying every field by an overall factor of $i$ should be a symmetry (this is in fact the gauge symmetry
corresponding to a rotation by $\pi$ around the x-axis).

Under this symmetry, we have
\bea
V^{1}=\left(\begin{array}{c}1\\
0\\0\\d\end{array}\right)
\rightarrow 
\left(\begin{array}{c}id\\
0\\0\\i\end{array}\right)
\eea 
We can further use a gauge transformation 
by
$e^{iL_3}$ (i.e. a gauge transform by
the $L_3$
subgroup of $SU(2)$) to transform
\bea
V^{I}=\left(\begin{array}{c} V^{I}_{111}\\V^{I}_{112}\\ V^{I}_{122}
\\ V^{I}_{222}\end{array}\right)
\to 
\left(\begin{array}{c} e^{3i\alpha}V^{I}_{111}\\e^{i\alpha}V^{I}_{112}\\ e^{-i\alpha}V^{I}_{122}
\\e^{-3i\alpha} V^{I}_{222}\end{array}\right)
\eea
A suitable choice of complex $\alpha$ allows us to bring 
\bea
V^{1}=\left(\begin{array}{c}1\\
0\\0\\d\end{array}\right)
\rightarrow 
\left(\begin{array}{c}id\\
0\\0\\i\end{array}\right)
\rightarrow 
\left(\begin{array}{c}1\\
0\\0\\-d\end{array}\right)
\eea 
We have therefore produced a configuration 
of the form (\ref{fieldform1}) 
but with a change in sign of $d$.
This implies that the sign of $d$ can be changed by 
a gauge transformation.
The gauge invariant combination is  $(d)^2$.

We now look for a $SU(2)$ invariant tensor 
such that gauge fixing the field to be of the form (\ref{fieldform1}) allows
us to deduce the value of $(d)^2$.

There is no nonzero bilinear 
invariant involving $V^1$ alone. We can 
however find an  invariant of degree 4 in $V^1$ 
as follows: we first construct
a symmetric combination of two fields
\bea
W^{IJ}_{ab}=(V^{I})_{acd}(V^{J})^{~cd}_{b}
+(V^{I})_{bcd}(V^{J})^{~cd}_{a}
\eea
(as always in $SU(2)$ , indices are raised 
and lowered by the epsilon tensor).
We can then construct the invariant
\bea
I_4^{IJKL}=W^{IJ}_{ab}W^{KLab}
\label{Inv2}
\eea

We now evaluate
\bea
I_4^{1111}=-8(d)^2
\eea

The knowledge of
the invariant $I_4^{1111}$ 
is therefore sufficient to deduce
the value of $d^2$, and therefore 
is 
hence sufficient to completely 
parametrize the gauge-inequivalent
configurations of a single field.
By the theorem cited in the introduction, 
$I_4^{IJKL}$ is a basis set of invariants for a single field in the dimension 4 representation of $SU(2)$.

\vskip 1 cm

We now consider multiple fields.
These can be brought by a complex 
gauge transformation to the form
\bea 
V^{1}=\left(\begin{array}{c}1\\
0
\\0
\\ d^{1}\end{array}\right)
\qquad \qquad 
V^{I}=\left(\begin{array}{c}a^{I}\\
b^{I}\\c^{I}\\d^{I}\end{array}\right)
\quad {\rm for\ I>1}
\label{firstconfig}
\eea

Exactly as above,
this parametrization  breaks the gauge symmetry but preserves a discrete $Z_2$
gauge symmetry.
Under this symmetry, we have
\bea 
V^{1}=\left(\begin{array}{c}1\\
0
\\0
\\ d^{1}\end{array}\right)\to 
\left(\begin{array}{c}1\\
0
\\0
\\ -d^{1}\end{array}\right)~~~~~~~~
\nonumber
\\
V^{I}=\left(\begin{array}{c}a^{I}\\
b^{I}\\c^{I}\\d^{I}\end{array}\right)
\to 
\left(\begin{array}{c} 
(d^1)^{-1}d^{I}
\\
-(d^1)^{-{1\over 3}}c^{I}
\\
(d^1)^{{1\over 3}}b^{I}
\\
-(d^1)a^{I}
\end{array}\right)
\eea

We now find combinations of the 
nonzero parameters which are gauge invariant under the residual $Z_2$.
It is convenient, and no more difficult, to find combinations which are invariant both under this $Z_2$ and the $L_3$ subgroup of $SU(2)$.

We first form combinations that
are invariant under the $L_3$
subgroup of $SU(2)$; these are
\bea
a^Id^J,  a^Ic^Jc^Kc^L, 
b^Ic^J, b^Ib^Jb^Kd^L
\nonumber
\eea
as well as expressions where one or
more of the indices is replaced by 1:
\bea
d^J,  c^Jc^Kc^L,
a^Id^1, b^Ib^Jb^Kd^1
\nonumber
\eea

Under the $Z_2$ action, these combinations are acted on as
\bea
d^1 \to  -d^{1}\qquad
a^{I}(d^1)\leftrightarrow  -d^I
~~~~~~~~~
\nonumber
\\
a^Id^J  \to -a^Jd^I\qquad
b^Ic^J \to -b^Jc^I ~~~~~~~
\\
a^Ic^Jc^Kc^L\leftrightarrow  b^Jb^Kb^Ld^I  \qquad 
b^Ib^Jb^Kd^1\leftrightarrow c^Ic^Jc^K
\nonumber
\eea

We make  combinations which are even/odd under the $Z_2$ gauge symmetry; the 
gauge invariant even combinations are
\bea
i_1^I&=&a^{I}(d^1)-d^I
\nonumber
\\
i_2^{IJ}&=&a^Id^J   -a^Jd^I
\nonumber
\\
i_3^{IJ}&=&b^Ic^J  -b^Jc^I
\label{invs1}
\\
i_4^{IJK}&=&b^Ib^Jb^Kd^1+ c^Ic^Jc^K
\nonumber\\
i_5^{IJKL}&=&a^Ic^Jc^Kc^L+ b^Jb^Kb^Ld^I
\nonumber
\eea

A product of two combinations odd under the $Z_2$
is even under the $Z_2$. We can therefore generate a new set of gauge invariant even combinations:
\bea
i_6^I&=&d^1(a^{I}d^1+d^I)
\nonumber
\\
i_7^{IJ}&=&d^1(a^Id^J + a^Jd^I)
\nonumber\\
i_8^{IJ}&=&d^1(b^Ic^J + b^Jc^I)
\label{invs2}
\\
i_9^{IJK}&=&d^1(b^Ib^Jb^Kd^1 - c^Ic^Jc^K)
\nonumber\\
i_{10}^{IJKL}&=&d^1(a^Ic^Jc^Kc^L- b^Jb^Kb^Ld^I)
\nonumber
\eea

The combinations $i_{1..10}$ parametrize the space of 
gauge inequivalent configurations.

We look for $SU(2)$ invariants that can 
reproduce these combinations; that is,  we 
look for $SU(2)$ invariant polynomials in 
the fields, such that when these are evaluated 
on the gauge fixed configuration (\ref{firstconfig}), their values are
sufficient to reconstruct the combinations (\ref{invs1}, \ref{invs2}).
It is immediate that a set of such operators would completely parametrize the configuration space.

The flavor symmetry is a guide. For instance,
the combinations with one free index
i.e. $i_1^I, i_6^I$
 should be reproduced from operators with 
 one free index. One such operator is 
provided by the operator $I_4^{IJKL}$ where
three indices are replaced by 1.
Another operator that we can consider is the antisymmetric bilinear
\bea
I_2^{IJ}= V^I_{abc}V^{Jabc}
\label{Inv1}
\eea

Indeed, we find that on the configuration (\ref{firstconfig})
\bea
I_2^{I1}=i_1^I
\qquad 
I_4^{111I}=4i_6^I
\eea
Hence a knowledge of the invariants 
(\ref{Inv2}, \ref{Inv1}) indeed allows us to  
reproduce the combinations with one free 
flavor index $i_1^I, i_6^I$.

The combinations with two free flavor
indices are $i_2^{IJ},i_3^{IJ},
i_7^{IJ},i_8^{IJ}$. We find that
on the configuration (\ref{firstconfig})
\bea
I_2^{IJ}=i_2^{IJ}-3i_3^{IJ}
\qquad 
I_4^{11IJ}=4i_7^{IJ}-4i_8^{IJ}
\eea
 but these are insufficient to reproduce
 $i_2^{IJ},i_3^{IJ},
i_7^{IJ},i_8^{IJ}$. We therefore
 need 
 further invariants of degree 6 and 8. 

A suitable choice are  the invariants
\bea
I_6^{IJKLMN}=W^{IJ}_{ab}W^{KL}_{cd}V^{Mab}_{~~~~~e}V^{Necd}~~~~~
\label{Inv3}
\\
I_8^{IJKLMNPQ}=W^{IJ}_{ab}W^{KL}_{cd}W^{MN}_{ef}V^{Pabc}V^{Qdef}
\label{Inv4}
\eea
We find on the configuration (\ref{firstconfig})
\bea
I_6^{1111IJ}=16(d^1)^2 i_3^{IJ}
 \\ 
 I_8^{111111IJ}=-32(d^1)^2i_8^{IJ}
\eea

Hence a knowledge of the invariants 
(\ref{Inv3}, \ref{Inv4}) allows us to  
reproduce the combinations with two free flavor
indices $i_2^{IJ},i_3^{IJ},
i_7^{IJ},i_8^{IJ}$.

We have two combinations with
three flavor indices i.e. $i_4^{IJK},
i_9^{IJK}$. 
We find on the configuration (\ref{firstconfig})
\bea
I_4^{1IJK}=-8i_4^{IJK}+..
\\
I_6^{111IJK}=-8i_9^{IJK}+..
\eea
where we have omitted terms which are composed
of products of invariants of lower degree.
So we do not
need further invariants to reproduce the combinations with three flavor indices.

When we move to the combinations with four indices, we find that
$i_{10}^{IJKL}$ cannot be reproduced by the invariants we have. We need a different invariant with six fields, which is
\bea
\tilde{I}_6^{IJKLMN}=W^{IJ}_{ac}W^{KL}_{bd}V^{Mab}_{~~~~~e}V^{Necd}
\label{Inv5}
\eea
We find that on the configuration (\ref{firstconfig})
\bea
I_6^{11I(JKL)}=48
i_{10}^{IJKL}-16(i_{10}^{JIKL}+i_{10}^{LJKI}+i^{KIJL})
\nonumber
\\
\tilde{I}_6^{1I(JKL)1}=-24i^{IJKL}+16(i^{JIKL}+i^{KIJL}+i^{LJKI})
\nonumber
\eea
 which allows us to solve for $i_{10}^{IJKL}$.
Finally, it is
straightforward to show that
the combination $i_5^{IJKL}$ with
four flavor indices  
can be reproduced
from $I_4^{IJKL}, I_8^{IJ1111KL}$.

We find then that the invariants
\bea
I_2^{IJ},I_4^{11IJ}, I_6^{IJKLMN},
\tilde{I}_6^{IJKLMN}, I_8^{IJKLMNPQ}
\eea
are sufficient to reconstruct the
gauge invariant parameter space of
this theory. The theorem from the introduction
then tells us that these are a complete set of 
independent polynomial invariants for the
dimension 4 representation of $SU(2)$.

\section{SO(3) with fields in the dimension-5 representation}

 We will take as our next example a 
 theory with a gauge group
 $SO(3)$ and a field content 
 where there are $N$ fields are in a representation of
dimension 5;
such a field is a 
symmetric tensor $V^{I}_{ij}$ of $SO(3)$, where
we take $i,j=1..3$, and $I=1..N$ is a flavor index 
labeling the different fields.
The field is a traceless symmetric tensor of $SO(3)$, 
and 
can therefore
be written as
\bea
V^{I}=\left(\begin{array}{ccc}V
^{I}_{11}&V^{I}_{12}&
V^{I}_{13}\\
V^{I}_{12}&V^{I}_{22}&V^{I}_{23}\\
V^{I}_{13}&V^{I}_{23}&V^{I}_{33}\end{array}
\right)
\eea
with $V^{I}_{11}+V^{I}_{22}+V^{I}_{33}=0$.
 
 It will prove convenient to define a product
\bea
(A\cdot B)_{ij}=A_{ik}B^{k}_{~j}
\eea
as well as a trace 
\bea
Tr(A)=A_{ij}\delta^{ij}
\eea
We may then write a sequence of invariants
\bea
I_2^{IJ}&=&Tr (V^{I}\cdot{V}^{J})\nonumber
\\
I_3^{IJK}&=&Tr(V^{I}\cdot{V}^{J}\cdot {V}^{K})\nonumber
\\
I_4^{IJKL}&=&Tr(V^{I}\cdot{V}^{J}\cdot {V}^{K}
\cdot {V}^{L})
\\
I_5^{IJKLM}&=&Tr(V^{I}\cdot{V}^{J}\cdot {V}^{K}
\cdot {V}^{L}\cdot {V}^{M})\nonumber
\label{invts}
\eea
and so on. 

To  find the independent set
of invariants, 
we now find the gauge-fixed configuration space, 
and attempt to 
characterize this space by invariants.

\vskip 1 cm

We first 
consider the case where the
matter content is a single
field $V^{1}$.
We can use complex gauge transformations to bring this to
the form of a diagonal matrix
\bea
V^{1}=\left(\begin{array}{ccc}a&0&0\\
0&b&0\\
0&0&c\end{array}\right)
\eea

This  generically
fixes the continuous gauge symmetry, 
but the ordering of the three diagonal elements
can be changed by a gauge transformation. There is therefore a residual discrete $Z_2\times Z_2$  gauge symmetry. 
In the special case when two eigenvalues coincide, the continuous symmetry is partially unbroken and there is a residual $U(1)$ symmetry.

The 
invariants are the symmetric combinations
\bea
i_1=a+b + c=0
\\
i_2=a^2+b^2+c^2
\\
i_3=a^3+b^3+c^3
\eea

We look for  $SO(3)$ invariants which can reproduce
 these combinations.
We find
\bea
I_2^{11}=i_2
\qquad
I_3^{111}=i_3
\eea

The invariants $I_2^{IJ},I_3^{IJK}$ are 
hence sufficient to completely 
parametrize the gauge-inequivalent
configurations of a single field,
and are hence form a complete set of invariants for one field.

\vskip 1 cm
We now consider
a generic configuration of multiple fields 
$V^{I}$.
We gauge fix $V^{1}$ as before.  The configuration is now
\bea
V^{1}=\left(\begin{array}{ccc}a&0&0\\
0&b&0\\
0&0&-a-b\end{array}\right)~~~~~~~~~~~~~~~~~
\nonumber
\\ 
V^{I}=\left(\begin{array}{ccc}
V^{I}_{11}&V^{I}_{12}&V^{I}_{13}\\
V^{I}_{21}&V^{I}_{22}&V^{I}_{23}\\
V^{I}_{31}&V^{I}_{32}&V^{I}_{33}\end{array}\right)\qquad {\rm for\ I>1}
\label{configs2}
\eea

\vskip 1 cm
We first consider the special case where two 
eigenvalues of $V^{1}$ coincide. Here we have that
\bea
V^{1}=a\left(\begin{array}{ccc}1&0&0\\
0&1&0\\
0&0&-2\end{array}\right)~~~~~~~~~~~~~~~~~
\nonumber
\\
V^{I}=\left(\begin{array}{ccc}
V^{I}_{11}&V^{I}_{12}&V^{I}_{13}\\
V^{I}_{21}&V^{I}_{22}&V^{I}_{23}\\
V^{I}_{31}&V^{I}_{32}&V^{I}_{33}\end{array}\right)\quad {\rm for\ I>1}
\eea

This particular configuration preserves a $U(1)$ 
subgroup of the $SU(2)$, so we could proceed by further fixing the gauge to find the
completely gauge-fixed hypersurface in field space. But the
gauge group is simple enough at this point that we can
straightforwardly write down a set of independent
polynomial combinations (invariant under the
$U(1)$ symmetry)
which parametrize the configuration space.

The fields in $V^{I}$
can be organized into combinations with 
specific charges $2, 1, 0, -1, -2$ under the $U(1)$; these are
\bea
 T^{I}_2 &=& 2V^{I}_{12}+i(V^{I}_{11}-V^{I}_{22})
 \nonumber\\
T^{I}_1 &=&
V^{I}_{13} - iV^{I}_{23}\\
 T^{I}_0 &=& V^{I}_{11} + V^{I}_{22},
\nonumber
\eea
where the subscripts denote the
respective charges. The negatively charged fields
are the complex conjugates of
the positive charges. 

In addition to the $U(1)$, there is also a
  discrete $Z_2$ symmetry
corresponding to 
charge conjugation
\bea
T^{I}_2\leftrightarrow -T^{I}_{-2},~~ 
T^{I}_1\leftrightarrow -T^{I}_{-1},~~T^{I}_0\leftrightarrow T^{I}_{0}
\eea

The combinations of fields invariant under the $U(1)$ symmetry are
\bea
 T^{I}_2 T^{J}_{-2},\
 T^{I}_1 T^{J}_{-1},\
T^{I}_0, \
\nonumber
T^{I}_2 T^{J}_{-1}T^{K}_{-1}, \
T^{I}_{-2} T^{J}_{1}T^{K}_{1}
\eea
Under the $Z_2$ action, these combinations are acted on as
\bea
T^{I}_0 &\leftrightarrow& T^{I}_0 
\nonumber
\\
 T^{I}_2 T^{J}_{-2}&\leftrightarrow&  T^{J}_2 T^{I}_{-2}
 \nonumber
 \\ 
 T^{I}_1 T^{J}_{-1}&\leftrightarrow&  T^{J}_1 T^{I}_{-1}
 \\
 T^{I}_2 T^{J}_{-1}T^{K}_{-1}
&\leftrightarrow &
-T^{I}_{-2} T^{J}_{1}T^{K}_{1}
\nonumber
\eea

We make  combinations which are even/odd under the $Z_2$; the gauge invariant 
even combinations are
\bea
i_1^{I} &\equiv& T^{I}_0
\nonumber
\\
i_2^{IJ} &\equiv& T^{I}_2 T^{J}_{-2}+T^{J}_2 T^{I}_{-2}
\nonumber
\\
i_3^{IJ} &\equiv& T^{I}_1 T^{J}_{-1}+T^{J}_1 T^{I}_{-1}
\label{invs4}
\\
i_4^{IJK} &\equiv& T^{I}_2 T^{J}_{-1}T^{K}_{-1}- T^{I}_{-2} T^{J}_{1}T^{K}_{1}
\nonumber
\eea
A product of two combinations odd under the $Z_2$
is even under the $Z_2$. The only such combination which cannot be written in terms of the already obtained even combinations is
\bea
i_5^{IJKL}\equiv (T^{I}_2 T^{J}_{-2}-T^{J}_2 T^{I}_{-2})
( T^{K}_1 T^{L}_{-1}-T^{L}_1 T^{K}_{-1})
\label{invs5}
\eea

The combinations 
$i_1^{I},
i_2^{IJ},
i_3^{IJ},
i_4^{IJK}, i_5^{IJKL}$
completely
parametrize the gauge-inequivalent orbits of the
configuration space.

\vskip 1 cm
We  now promote these to
$SO(3)$ invariants; that is, we look for $SO(3)$ invariants
which reduce to  
the  combinations
(\ref {invs4},\ref {invs5}) on the gauge-fixed configuration space
of equation (\ref{configs2}).
Once again, we use the flavor symmetry as a guide.

The combination with one free index
i.e. $i_1^I$
 should be reproduced from operators with 
 one free flavor index. One such operator is 
provided by the operator $I_2^{IJ}$ where
one of
the fields is taken to be $V^1$. 
Indeed, we find
\bea
I_2^{1I}=3a  i_1^{I}
\eea
Hence a knowledge of the invariant 
$I_2^{IJ}$ allows us to  
reproduce the combinations with 
one free flavor index $i_1^I$.

The combinations with two free flavor indices i.e. $i_2^{IJ}, i_3^{IJ}$ should be reproduced from 
$SO(3)$ invariants with two free flavor indices $IJ$. 
Indeed, we find 
\bea
I_2^{IJ}
= {1\over 4}i_2^{IJ}+ i_3^{IJ}+...
\\
I_3^{1IJ}
= {a\over 4} (i_2^{IJ}-2i_3^{IJ})+...
\eea
where the ellipses indicate terms with (already determined) lower degree combinations like $i_1^{I}i_1^{J}$.
These two invariants therefore determine $i_2^{IJ}, i_3^{IJ}$.

For 
$i_4^{IJK}$, we have $JK$ symmetrized. This will
be reproduced by an $SO(3)$ invariant with three 
flavor indices $IJK$, where the $JK$ indices are symmetrized. 
Such a $SO(3)$ invariant, in general, 
when expanded in terms of the fields, would produce linear
combinations of 
$i_4^{IJK}+i_4^{JIK}+i_4^{KIJ}$ 
and $-i_4^{IJK}+i_4^{JIK}+i_4^{KIJ}$,
both of which satisfy these symmetries.
We therefore need at least { two} different  $SO(3)$ invariants with this
symmetry structure.

There is one such invariant
of degree 3 which is completely symmetric
\bea
I_3^{IJK}
=-{i\over 4}
(i_4^{IJK}+i_4^{JIK}+i_4^{KIJ})
\eea
where the ellipses indicate terms with (already determined) lower degree combinations. We need at least one more invariant, and so we now consider invariants of degree 4.

A consideration of the combination $i_5^{IJKL}$
suggests that we should look at combinations where the four indices form two antisymmetrized pairs.
We choose such a combination 
\bea
\tilde{I}_4^{IJKL}&=&
Tr(V^{[I}\cdot{V}^{J]}\cdot {V}^{[K}
\cdot{V}^{L]})
\eea
and we find
\bea
\tilde{I}_4^{IJK1}+\tilde{I}_4^{IKJ1}=
-(\frac{3ai}{2})(2i_4^{IJK}-i_4^{JIK}-
i_4^{KIJ})
\eea
Hence  the invariants $I_3^{IJK}, \tilde{I}_4^{IJKL}$ allow us to solve for $i_4^{IJK}$.

Finally we consider 
$i_5^{IJKL}$. Here $IJ$ and $KL$ are antisymmetrized. We should look for 
$SO(3)$ invariants with four 
flavor indices $IJKL$, where the  
$IJ, KL$ indices are antisymmetrized. 
However, such an $SO(3)$ invariant, in general, 
when expanded in terms of the fields, will produce linear
combinations of  

(a) the 
combination completely antisymmetric
in $IJKL$:
 \bea
 i_5^{IJKL}+i_5^{KLIJ}-i_5^{IKJL}+i_5^{JKIL}+ 
i_5^{ILJK}-i_5^{JLIK}\nonumber
\eea
(b) another combination, still symmetric in $(IJ)\leftrightarrow (KL)$:
\bea
 i_5^{IJKL}+i_5^{KLIJ}+(1/2)
 (i_5^{IKJL}-i_5^{JKIL}- 
i_5^{ILJK}+i_5^{JLIK})\nonumber
\eea
 and (c) one combination antisymmetric in $(IJ)\leftrightarrow (KL)$
 \bea
 i_5^{IJKL}-i_5^{KLIJ}\nonumber
 \eea

We should therefore have  at least {\it three}  $SO(3)$ invariants
with   four flavor indices $IJKL$, where the  
$IJ, KL$ indices are antisymmetrized. 
One such invariant is provided by $\tilde{I}_4^{IJKL}$.
We therefore
need two invariants of degree 5.
%

We therefore define
\bea
\tilde{I}_5^{IJKLM}=I_5^{[IJKLM]}
\\
{I}_6^{IJLKM}=I_5^{[IJ][KLM]}
\eea


We find that
$\tilde{I}_5^{IJKL1}$ is proportional to the
completely antisymmetric combination (a),
 $\tilde{I}_4^{IJKL}$ is proportional to the
combination (b), and
$ 
I_5^{[IJ1][KL]} 
-I_5^{[KL1][IJ]}$ is proportional to 
combination (c).
Hence
$i_5^{IJKL}$ can indeed be written as 
a combination of these
invariants.

The configuration space with the enhanced $U(1)$ symmetry is therefore completely
characterized by the
invariants
\bea
I_2^{IJ},
I_3^{IJK}, 
\tilde{I}_4^{IJKL},
\tilde{I}_5^{IJKLM},
{I}_6^{IJLKM}
\nonumber
\eea

\vskip 1 cm

More generally, the eigenvalues of $V^1$ are all different, and we have
\bea
V^{1}=\left(\begin{array}{ccc}a&0&0\\
0&b&0\\
0&0&-a-b\end{array}\right)~~~~~~~~~~~~~~~
\nonumber
\\
V^{I}=\left(\begin{array}{ccc}
V^{I}_{11}&V^{I}_{12}&V^{I}_{13}\\
V^{I}_{21}&V^{I}_{22}&V^{I}_{23}\\
V^{I}_{31}&V^{I}_{32}&V^{I}_{33}\end{array}\right)\quad {\rm for\ I>1}
\label{configs3}
\eea

The parameters $a,b$ are reproduced from 
$I_2^{11},I_3^{111}$, following the analysis for a single field.

This parametrization
of the configuration space  preserves 
two $Z_2$ symmetries:
 \bea
 (Z_2)_A: V^{I}_{12}\rightarrow -V^{I}_{12}, V^{I}_{13}\rightarrow -V^{I}_{13}
 \nonumber
 \\
  (Z_2)_B: V^{I}_{12}\rightarrow -V^{I}_{12}, V^{I}_{23}\rightarrow -V^{I}_{23}
 \eea

To find the moduli space, we should find the
 combinations of $V^{I}_{ij}$ which are gauge invariant 
 under these discrete symmetries. Once again, the remaining  symmetry
is simple enough that we can just do this by inspection.
We find that the polynomials invariant under the two
discrete symmetries are generated by
\bea
V^{I}_{11},~~V^{I}_{22},~~
V^{I}_{12}V^{J}_{12},~~
V^{I}_{13}V^{J}_{13}, ~~
V^{I}_{23}V^{J}_{23},~~
V^{I}_{12}V^{J}_{13}V^{K}_{23}\nonumber
\eea

We  now promote these to
$SO(3)$ invariants; that is, we look for $SO(3)$ invariants
which reduce to  
these  combinations
 on the configuration space
of equation (\ref{configs3}). We first check whether the invariants we
have found are
sufficient to do this.

We start with invariants with one flavor index $I$. We find
\bea
I_2^{1I}=aV^{I}_{11}+bV^{I}_{22}
+(a+b)(V^{I}_{11}+V^{I}_{22})
\\
I_3^{11I}=a^2V^{I}_{11}+b^2V^{I}_{22}
+(a+b)^2(-V^{I}_{11}-V^{I}_{22})
\eea
which can be used to solve for 
$V^{I}_{11}, V^{I}_{22}$ in terms
of $I_2^{1I}, I_2^{11I}$.
This inversion fails only if
$(b+2a)=0, (2b+a)=0$ or $a=b$ i.e.
when two eigenvalues in $V^{1}$ are equal, which we have already assumed to not be the case.

Similarly, it
is easy to show that the invariants $I_2^{IJ}, I_3^{IJ1}$ are
sufficient to  reconstruct
$V^{I}_{12}V^{J}_{12}, V^{I}_{13}V^{J}_{13}, V^{I}_{23}V^{J}_{23}$,
and $I_3^{IJK}, \tilde{I}_4^{IJK1}, {I}_6^{IJ1K1}$ are
sufficient to solve for $V^{I}_{12}V^{J}_{13}V^{K}_{23}$.

Our final result is then that
every point on the gauge invariant configuration space can be reproduced by a 
knowledge of the invariants
\bea
I_2^{IJ},\
I_3^{IJK},\ 
\tilde{I}_4^{IJKL},\
\tilde{I}_5^{IJKLM}
,\
{I}_6^{IJLKM}
\eea
Hence the theorem described in the introduction
ensures that any gauge invariant polynomial in this theory  can be generated by these invariants.

 \section{SU(3) with sextets}
 
 We now consider a theory with a $SU(3)$ 
 symmetry and fields in the sextet
 representation $V^{I}_{ij}$.
 
 We can form an infinite set of
 invariants by contracting these
 fields with the epsilon tensor $\epsilon^{ijk}$. We now
 find an independent set of tensors by finding
 a set that can parametrize the gauge fixed configuration space.

 \vskip 1 cm

We begin by considering a single field. By a gauge transformation, we can bring it to the form
 \bea
V^{1}=\left(\begin{array}{ccc}a&0&0\\
0&b&0\\
0&0&c\end{array}\right)
 \eea
 
 This form of the field preserves
 two $U(1)$ symmetries; the first is
 \bea
 V_{11}\to e^{2i\alpha}V_{11}\qquad 
V_{22}\to e^{-2i\alpha}V_{22} \qquad   V_{33}\to  V_{33}
\nonumber
\\
 V_{13}\to e^{i\alpha}V_{13}\qquad 
V_{23}\to e^{-i\alpha}V_{23}
\qquad 
 V_{12}\to  V_{12} 
 \eea
 and the second is
  \bea
 V_{11}\to e^{2i\alpha}V_{11}\qquad 
V_{33}\to e^{-2i\alpha}V_{33} \qquad   V_{22}\to  V_{22}
\nonumber
\\
 V_{12}\to e^{i\alpha}V_{12}\qquad 
V_{23}\to e^{-i\alpha}V_{23}
\qquad 
 V_{13}\to  V_{13} 
 \eea
 
 These symmetries alter the eigenvalues without changing the form; the first one takes
 $a\to e^{2i\alpha}a, b\to e^{-2i\alpha}b, c\to c$, and the second takes
  $a\to e^{2i\alpha}a, c\to e^{-2i\alpha}c, b\to b$.
 The only gauge invariant combination is the product $abc$.
 
We define the invariant
(we use dotted indices for the complex conjugate representation)
\bea
I_3^{IJK}\equiv \epsilon^{\dot{i}\dot{k}\dot{m}}
\epsilon^{\dot{j}\dot{l}\dot{n}}V^{I}_{ij}V^{J}_{kl}V^{K}_{mn}
 \eea
 and we find
 \bea
 I_3^{111}=6abc
 \eea
 This invariant therefore reproduces the 
 gauge-fixed configuration space for a single field, and is therefore a complete set of invariants for a single sextet of $SU(3)$.
 
  \vskip 1 cm

We now consider multiple fields,
 We can bring these to the form
  \bea
V^{1}=\left(\begin{array}{ccc}a&0&0\\
0&a&0\\
0&0&a\end{array}\right)~~~~~~~~~~~~~~~
~~~~~~~
\nonumber
\\
V^{I}=\left(\begin{array}{ccc}V^{I}_{11}&V^{I}_{12}&V^{I}_{13}\\
V^{I}_{12}&V^{I}_{22}&V^{I}_{23}\\
V^{I}_{13}&V^{I}_{23}&V^{I}_{33}\end{array}\right) \quad {\rm for\ I>1}
\label{configs7}
 \eea
 where we have used the $U(1)$ symmetries to further simplify the form of the first field.
 
The form of the configuration space still preserves an $SO(3)$ symmetry, and each sextet of
$SU(3)$ decomposes to a 1+5 of $SO(3)$. 
Fortunately, we have analyzed this
system already in the previous section, and so we can write down the
invariants. The only new invariant is the 
singlet, which is the trace of the matrix. Combining with
the previously derived $SO(3)$ invariants for a dimension 5 field,
the $SO(3)$ invariant combinations for this theory are
\bea
i_1^I &\equiv& Tr(V^I)\nonumber
\\
i_2^{IJ} &\equiv& Tr ((V^{I}\cdot{V}^{J})\nonumber
\\
i_3^{IJK} &\equiv& Tr(V^{I}\cdot{V}^{J}\cdot {V}^{K})
\nonumber
\\
\tilde{i}_4^{IJKL} &\equiv& Tr(V^{[I}\cdot{V}^{J]}\cdot {V}^{[K}
\cdot {V}^{L]})
\\
\tilde{i}_5^{IJKLM} &\equiv& Tr(V^{[I}\cdot{V}^{J}\cdot {V}^{K}
\cdot {V}^{L}\cdot {V}^{M]})
\nonumber
\\
{i}_6^{IJLKM} &\equiv& Tr(V^{[I}\cdot{V}^{J}\cdot {V}^{K]}
\cdot {V}^{[L}\cdot {V}^{M]})
\nonumber
\eea

We now find $SU(3)$ invariants which reproduce these 
combinations on the configuration space 
(\ref{configs7}).

From the invariant that we have already defined, we obtain 
\bea
I_3^{11I}&=&2a^2i_1^I\nonumber
\\ 
I_3^{1IJ}&=&-a i_2^{IJ}+..
\\
I_3^{IJK}&=&2i_3^{IJK}+..\nonumber
\eea
which reproduces all combinations with one, two or three free flavor indices. 

For the remaining combinations, we need to consider
invariants containing six fields contracted with 4 epsilon tensors. The structure of the combinations above suggests
that we should look at combinations where there are two
pairs of three fields, where the three fields are antisymmetrized.
This suggests  the invariant
\bea
I_6^{IJKLMN}=\epsilon^{\dot{a}\dot{b}\dot{c}}
V^I_{ad}
V^J_{be}
V^K_{cf}\
\epsilon^{\dot{e}\dot{i}\dot{m}}\epsilon^{\dot{f}\dot{k}\dot{n}}
\epsilon^{\dot{d}\dot{j}\dot{l}}
V^L_{ij}V^M_{kl}V^N_{mn}
\eea

We find
\bea
I_6^{[IJ]1[KL]1}=-a^2\tilde{i}_4^{IJKL}
\\
I_6^{[IJKLM]1}=6ai_5^{IJKLM}
\\
I_6^{[IJK][LM]1}=4ai_6^{IJKLM}
\eea

We thus find that
every point on the gauge-fixed  configuration space can be reproduced by a 
knowledge of the invariants
\bea
I_3^{IJK}, I_6^{IJKLMN}
\eea
Hence the theorem described in the introduction
ensures that any gauge invariant polynomial in this theory  can be generated by these invariants.

\section{Summary and Conclusion}

We have discussed a new method to 
efficiently find a set of independent invariant tensors in gauge theories.
We have done this by using a theorem, familiar from supersymmetric field theories, that relates D-flat directions to the 
invariants in a gauge theory.
Specifically, this theorem asserts that the
constant configurations, identified by 
complex gauge transformations, are in 1-1 correspondence with the gauge invariant operators in the theory. 

We have shown that this provides a  straightforward method
 to find the independent invariant tensors.
We have explicitly applied these methods to three gauge theories-- $SO(3)$ with fields in the symmetric tensor representation, $SU(2)$ with a dimension 4 representation, and $SU(3)$ with matter in the sextet-- and in each case,
we have found the set of independent polynomial invariants. This shows the practicality of the approach.

Our methods are general, and as far as we can see, can be applied to 
any group with any matter content.
 The immediate ones which would be interesting to analyze are exceptional groups  with matter in the (anti)-fundamental representation.
Knowing the invariant tensors would also help in looking
for dual pairs in supersymmetric gauge theories with 
exceptional gauge groups.

We hope to return to this topic in future work.

\section*{Acknowledgments}

This work was supported in part by NSF Grant No.~PHY-1915005. The research of Y.A. was supported by Kuwait University.

\end{document}